\documentstyle[twoside,fleqn,espcrc2,epsfig]{article}

\newcommand{\AmS}{{\protect\the\textfont2
  A\kern-.1667em\lower.5ex\hbox{M}\kern-.125emS}}

\title{Block Spin Effective Action for Polyakov Loops in 4D SU(2) LGT
        \thanks{Poster presented by S. Vinti, preprint MS-TPI-97-9}}
\author{K. Pinn and S. Vinti\address{Institut f\"ur Theoretische Physik I,
        Universit\"at M\"unster, \\
        Wilhelm-Klemm-Str.~9, D-48149 M\"unster, Germany}}
       
\begin{document}

\begin{abstract}

Using a variant of the IMCRG method of Gupta and Cordery, we explicitly
compute majority rule  block spin effective actions for the signs of the
Polyakov loops in 4D SU(2) finite temperature lattice gauge theories. To
the best of our knowledge, this is the first attempt to compute
numerically effective actions for the Polyakov loop degrees of freedom
in 4D SU(2). The most important observations are: 1.~The
renormalization group flow at the deconfinement transition can be nicely
matched with the flow of the 3D Ising model, thus confirming the
Svetitsky-Yaffe conjecture. 2.~The IMCRG simulations of the FT SU(2)
model have strongly reduced critical slowing down.                      

\end{abstract}

\maketitle

\section{INTRODUCTION}

D+1-dimensional pure LGT's with finite periodic extension $N_t$ in the
``time'' direction describe glue systems at finite temperature
$1/N_t$. They undergo a deconfinement transition, signalled  by a
finite expectation value of the trace of the Polyakov loop.

According to the 15 years old Svetitsky-Yaffe conjecture~\cite{sy}, the
D-dimensional effective statistical model for the Polyakov loops has
an action (Hamiltonian) with short range interactions  and  a global
symmetry group given by  the center of SU(N).
If both the deconfining phase transition of the gauge
model and of the corresponding order-disorder phase transition of the
spin system are continuous, the two models belong to the same
universality  class.

The conjecture that the deconfinement transition of 4D SU(2) LGT
belongs to the 3D  Ising universality class is supported by analytical
calculations,  and numerical 
estimates of its critical indices, c.f.~\cite{Bielefeld}.

We tested the Svetitsky-Yaffe conjecture by comparing flows of block
spin effective actions for the Polyakov loops  with flows of the 3D
Ising model. Approach of both flows to a single trajectory 
ending in a common renormalization group fixed point  
demonstrates universality on a fundamental level.

\section{BLOCKING POLYAKOV LOOPS}

Our procedure to generate and analyse block spin effective actions
for Polyakov loops consists of the following steps.
\begin{itemize}
\item Map the SU(2) configurations $U$ living on an 
$N_t \times N_s^3$ 
lattice to Ising configurations $\sigma(U)$ on an $N_s^3$ lattice.
The Ising variables are 
given by the signs of the Polyakov loops. 
\item Block the Ising configurations with the majority rule, using
      cubical blocks of size $L_B$.
\item Compute the effective coupling constants using IMCRG~\cite{gupta}.
\item Generate a renormalization group flow by increasing the block 
      size $L_B$. 
\item Compare the resulting flow with that computed directly in the 
      Ising model. 
\end{itemize}

More formally, the effective action for the signs of the Polyakov loops
is given by 
\begin{equation}
\exp[-H'(\mu)]= \int DU~ P(\mu,U)~\exp[-S_{\rm g}(U)] \, , 
\end{equation}
with
\begin{equation}
P(\mu,U)
= \prod_{x'}^{\rm (blocks)} \frac12 \left[ 1 + \mu_{x'} \;
\mbox{sign} \sum_{x\in x'} \sigma_x(U) \right].
\end{equation}
Here, $S_{\rm g}$ is the standard Wilson action for SU(2).
In case of an even block size $L_B$ the sum of the Ising spins $\sigma_x$
inside a block $x'$ can be zero. In that case a positive (negative)
$\mu_{x'}$ is selected with probability one half.

Unfortunately, effective Hamiltonians contain an infinite number
of coupling constants. In practical calculations one has to
truncate to a finite set of interactions. We chose to include in 
the ansatz eight 2-point couplings and six 4-point couplings. 
The 2-point couplings can be labelled by specifying the relative
position of the interacting spins (up to obvious symmetries): 
our couplings $K_1 \dots K_8$ then correspond to
001,011,111,002,012,112,022,122. The 4-point couplings 
$K_9 \dots K_{14}$ are defined through Fig.~1. The 
corresponding interaction terms in the effective Hamiltonian
are denoted by $S_\alpha'$, $\alpha=1 \dots 14$.

\vskip5mm
\begin{center}
{~ }\epsfig{file=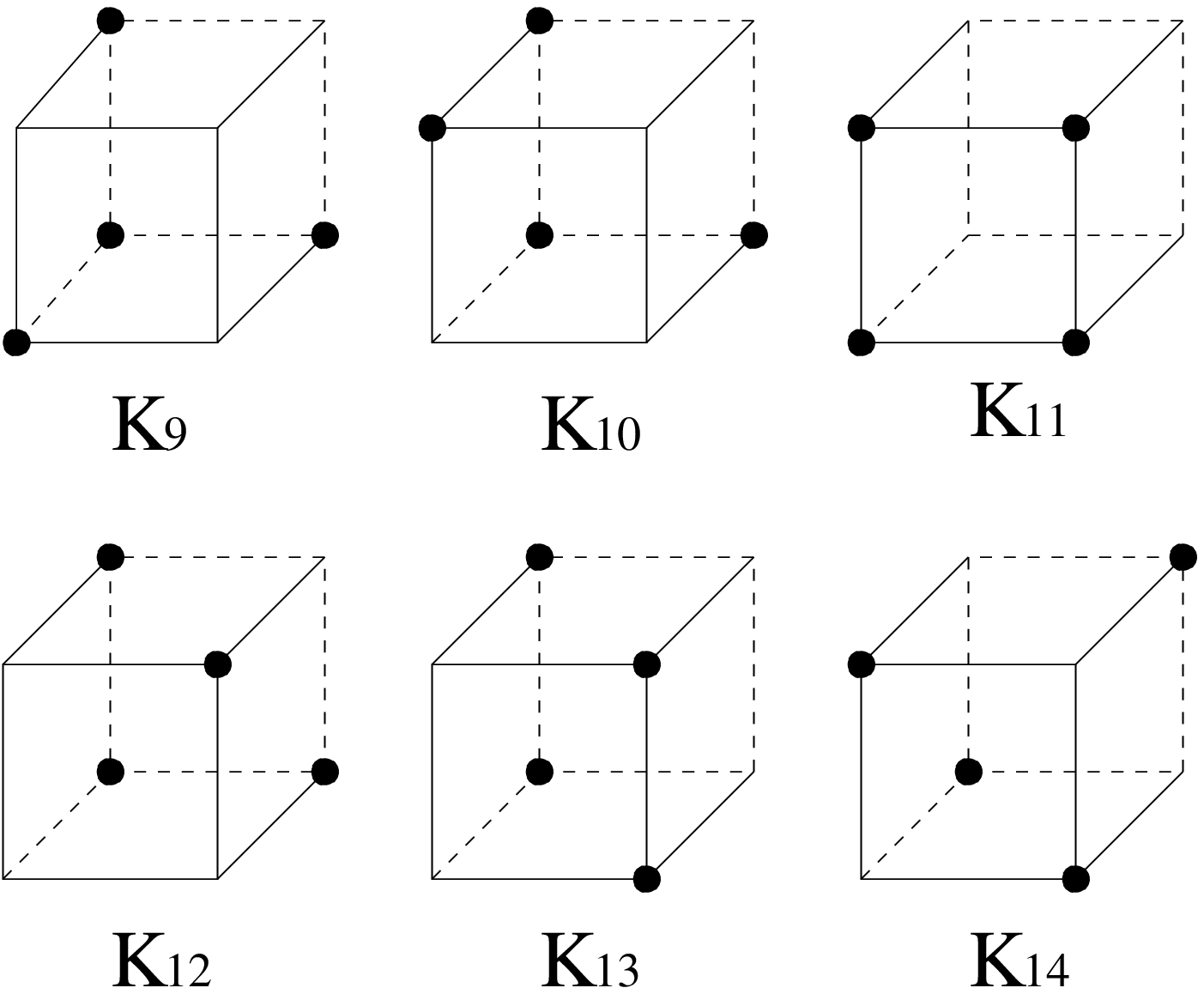,height=5.0cm}
\end{center}
Fig. 1. Definition of the 4-point-couplings.

\section{IMCRG FOR POLYAKOV LOOPS}

Translating the rationale of IMCRG~\cite{gupta} to the present context
means that one does {\em not} simulate standard SU(2) model but
instead the system with partition function 
\begin{equation}
\sum_{\mu} \int DU~ P(\mu,U)
\, \exp\left[ - S_{\rm g}(U)\; + \; \bar H(\mu)\right] \, , 
\end{equation}
where $\bar H(\mu)= \sum_{\alpha} \bar K_\alpha S_{\alpha}'(\mu)$
is a guess for $H'(\mu)$.
Note the {\em plus} sign in front of $\bar H$.
As before, $\mu$ denote the block spins, defined on blocks of size $L_B$
obtained from blocking the Polyakov loops.
The crucial observation is now that if $\bar H=H'$, i.e., if the
guess of the Hamiltonian is right, then the $\mu_{x'}$ decouple 
completely. This means in particular that the correlations
$\langle S_{\alpha}'(\mu)\rangle$ vanish. 
A non-perfect guess can be improved by iteration of 
\begin{equation}
\bar K_{\alpha} \longrightarrow \bar K_{\alpha} + n_\alpha^{-1}
\langle S'_\alpha(\mu) \rangle \, ,
\end{equation}
where  $n_\alpha$ are trivial multiplicity factors.
A few iterations of this correction step usually suffice to
obtain good precision for the effective couplings.

This method, though it seems to be restricted to Ising type models, has
several merits. One of them is the following: For small enough blocks,
the  autocorrelations in the simulations are drastically reduced. This 
follows from the fact that in case of perfect compensation the blocks
completely decouple and fluctuate independently.
In the SU(2) IMCRG calculations discussed below this phenomenon 
was clearly observed~\cite{tocome}.

\section{MONTE CARLO RESULTS}

We started by computing the flow of the critical 3D Ising
model, using two different models. The standard Ising 
model with nearest neighbour couplings becomes critical
at $\beta=0.2216544$. A version ``$I_3$'' which includes
also third (cube-diagonal) neighbour couplings has a critical point
at $\beta_1 = 0.128003$ and $\beta_3=0.051201$~\cite{bloete}. 

We then turned to FT SU(2) with $N_t=2$. We computed
the effective actions from simulations on lattices consisting of $6^3$ blocks
of size $L_B \leq 6$. The simulations were performed 
at $\beta_{\rm g}=1.880,1.877,1.874,1.871$. 
Best matching with the Ising data was achieved at
$\beta_{\rm g}=1.877$. 
The flow of $K_1$ and $K_2$ for this gauge coupling
is shown in Fig.~2, together with the
Ising flows. The gauge data are displayed with squares, 
the Ising results are shown with bars, diamonds and
dotted fit lines. 
The fits were done with a simple power law~\cite{tocome}.
A nice matching is observed also for the 12 couplings not
shown here. Note that the block sizes of the various
models have to be rescaled with respect to each other in order
to yield matching. E.g., the $L_B$ from the $I_3$ model
have to be rescaled by a factor of 0.59 with respect to
that of the standard Ising model.

Let us finally show a comparison of the flow of the
nearest neighbour coupling for different gauge
couplings. This is shown in Fig.~3. The dashed
line, together with diamonds and crosses, gives the
Ising flow. The other data (from top to 
bottom) correspond to $\beta_{\rm g}= 1.880$ (triangles),
1.877 (crosses), 1.774 (squares), and 1.771 (stars).

The gauge block sizes are rescaled such that best matching
is obtained for $\beta_{\rm g}= 1.877$. Within the given
precision, also the $\beta_{\rm g}=1.874$ data could
be rescaled to match the Ising flow. This is, however, clearly
ruled out for $\beta_{\rm g}=1.880$.

\vskip2mm
\epsfig{file=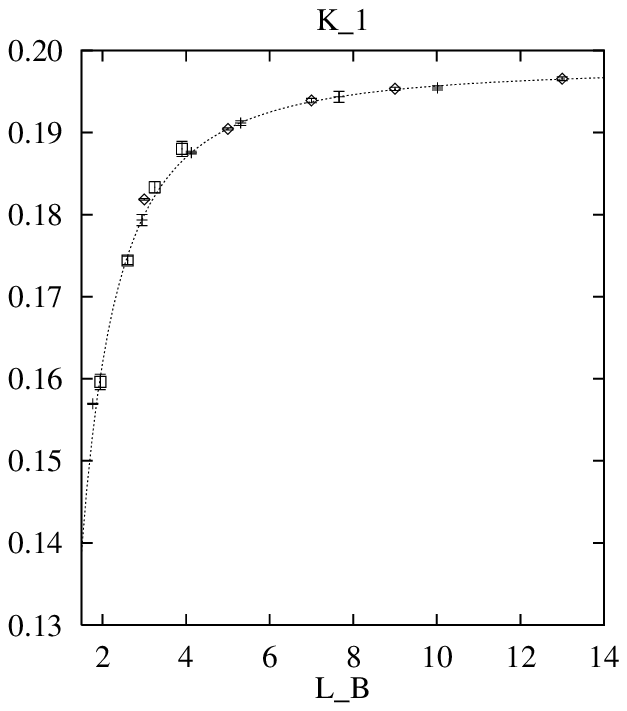,height=7.5cm}
\vskip0.2cm
\epsfig{file=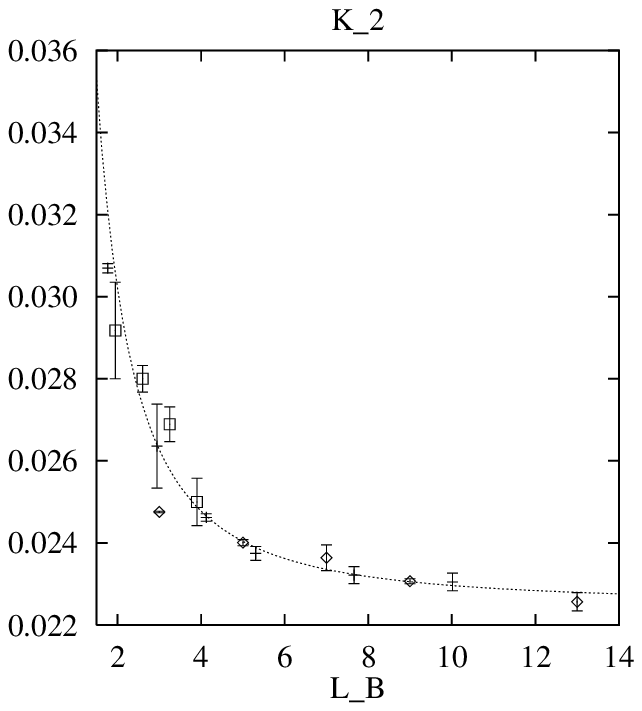,height=7.5cm}
Fig.~2. Matching of the couplings $K_1$ and $K_2$.
\vskip2mm

\vskip0.2cm
\epsfig{file=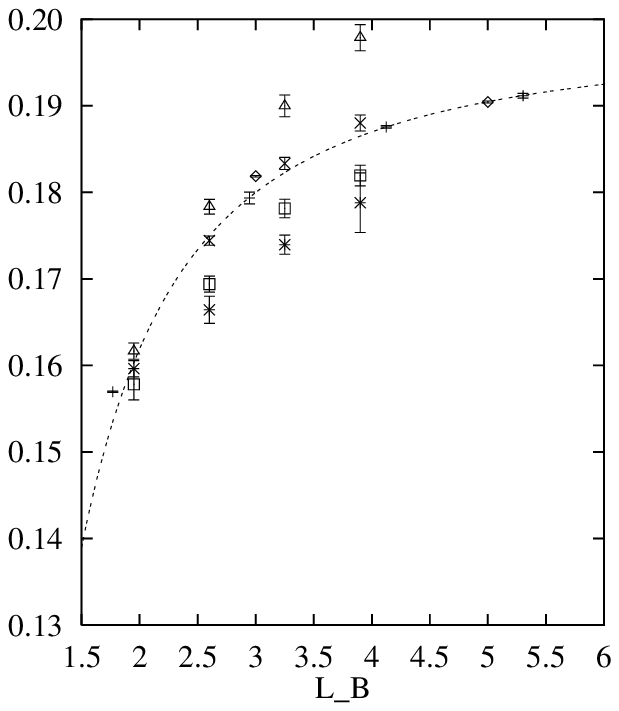,height=7.0cm}
Fig.~3. Comparison of $K_1$ flow for different $\beta_{\rm g}$.
\vskip0.2cm
                                        
\section{CONCLUSIONS}

IMCRG works well as a method to compute the effective action of Ising
type degrees of freedom, also in non-Ising models like 4D FT SU(2)
LGT. The calculations could be done on workstations. The Svetitsky-Yaffe
conjecture is confirmed in a very fundamental way by observing matching
of the RG trajectories with those of the Ising model. We obtained results for
$N_t=1$ also, see~\cite{tocome}. Extension to $N_t$ bigger than two is
expensive, because one needs larger blocks  to come close enough to the
fixed point.

\end{document}